\documentclass[9pt, conference]{IEEEtran}
\IEEEoverridecommandlockouts
\usepackage{cite}
\usepackage{amsmath,amssymb,amsfonts}
\usepackage{algorithmic}
\usepackage{graphicx}
\usepackage{textcomp}
\usepackage{svg}

\newcommand{\Zb}{\mathbf{Z}}
\newcommand{\Wb}{\mathbf{W}}
\newcommand{\bb}{\mathbf{b}}
\usepackage{url}
\usepackage{comment}
\usepackage{amssymb}
\usepackage{pifont}
\usepackage[hidelinks]{hyperref}
\newcommand{\cmark}{\ding{51}}%
\newcommand{\xmark}{\ding{55}}%
\def\BibTeX{{\rm B\kern-.05em{\sc i\kern-.025em b}\kern-.08em
    T\kern-.1667em\lower.7ex\hbox{E}\kern-.125emX}}
\begin{document}

\title{Learning Speaker-Activity-Conditioned WavLM Speaker Embeddings with Real Conversational Data
}
\author{
\IEEEauthorblockN{
Prachi Singh\IEEEauthorrefmark{1},
Nguyen Hai Phong\IEEEauthorrefmark{1},
Petr Pálka\IEEEauthorrefmark{1},
Jiangyu Han\IEEEauthorrefmark{1},\\
Marc Delcroix\IEEEauthorrefmark{2},
Shota Horiguchi\IEEEauthorrefmark{2},
Naohiro Tawara\IEEEauthorrefmark{2},
Lukáš Burget\IEEEauthorrefmark{1}
}
\IEEEauthorblockA{
\IEEEauthorrefmark{1}Brno University of Technology,	Czechia\\
\IEEEauthorrefmark{2}NTT, Inc.,	Japan\\
Email: prachi@fit.vut.cz, xnguye28@vutbr.cz
}
\thanks{The work was supported by Ministry of Education, Youth and Sports of the Czech Republic (MoE) through the OP JAK project  (ID:CZ.02.01.01/00/23\_020/0008518). Computing on IT4I supercomputer was supported by MoE through the e-INFRA CZ (ID:90254).}
}

\maketitle


\begin{abstract}
End-to-end neural diarization with vector clustering (EEND-VC) is a strong framework for long-form conversational diarization, where speaker embeddings are used to link local speaker decisions across windows. Conventional EEND-VC systems extract these embeddings by cutting out regions assigned to each local speaker, preferably using only non-overlapped speech. This works well when enough clean target-speaker speech is available, but it discards useful speaker evidence in overlapped regions and becomes unreliable when a local speaker appears mostly in overlap. We propose a speaker-activity-conditioned embedding extractor that adapts a pretrained WavLM-MHFA speaker verification model to extract target-speaker embeddings directly from full conversational windows. Inspired by the recently proposed diarization-conditioned Whisper (DiCoW) model for target-speaker ASR, frame-level diarization-dependent transformations inject local speaker-activity information into intermediate WavLM representations while preserving the pretrained initialization. Experiments on multiple conversational diarization benchmarks show that fine-tuning with limited real meeting data reduces speaker-confusion errors and improves EEND-VC diarization performance. The results further indicate that real conversational data provides an effective alternative to large-scale simulated-mixture training for diarization-aware speaker embedding extraction.
\end{abstract}


\section{Introduction}
\label{sec:intro}
Speaker diarization, the task of determining who spoke when, remains challenging for long-form conversational recordings with overlapping speech, short speaker turns, background noise, and an unknown number of speakers. A successful family of systems for such recordings is end-to-end neural diarization with vector clustering (EEND-VC)~\cite{bredin23_interspeech,han2025leveraging,kinoshita2021integrating}. In this framework, a local EEND model estimates frame-level speaker activity in short, typically overlapping windows. The speaker identities predicted in each window are only local, so speaker embeddings are extracted for the local speakers and clustered to assign consistent global speaker identities across the whole recording. Recent progress in local EEND models~\cite{han2025fine,hanpruning26,plaquet2025mamba} and clustering backends~\cite{landini2022bayesian,palka2026vbx,delcroix2023multi,palka2026SphereVBx} has improved EEND-VC diarization, but reliable speaker embeddings remain essential for linking local speakers across windows. Poor embeddings lead directly to speaker-confusion errors in the final diarization output.

Most speaker embedding extractors used in EEND-VC are trained for speaker verification on segments that typically contain speech from a single speaker~\cite{zeinali2019rvector,desplanques20_interspeech_ecapa,peng2023attention,wang2023cam++}. EEND-VC systems therefore usually construct the embedding-extractor input to satisfy this single-speaker assumption. For each local speaker, the system cuts out and concatenates subsegments assigned to that speaker by the local EEND output, preferably using only non-overlapped speech when available and falling back to overlapped regions when no clean speech exists. This is practical and effective when enough clean speech is available for each local speaker. However, in conversational recordings, some speakers may have little or no non-overlapped speech within a local window. In such cases, the system either ignores useful target-speaker evidence present in overlapped speech or must extract embeddings from speech mixed with interfering speakers.

Recent work has started to address this limitation by conditioning speaker embedding extraction on diarization information~\cite{horiguchi2025guided,horiguchi2025mitigating,kalda2025diarization}. These approaches show that speaker activity can help extract more useful embeddings for EEND-VC, especially when clean single-speaker regions are short or unavailable. However, prior systems typically train dedicated embedding models on simulated multi-speaker mixtures generated from VoxCeleb-style data~\cite{nagrani2020voxceleb}. This raises two practical questions: whether a strong pretrained speaker embedding extractor can be adapted without training a new diarization-conditioned model from scratch, and whether such adaptation can be done using a limited amount of real conversational data. Training on real meetings may expose the model to conversational effects that are difficult to reproduce faithfully in simulation.

In this paper, we propose a speaker-activity-conditioned embedding extractor for EEND-VC. Like prior diarization-conditioned approaches, the goal is to extract embeddings from the full local window rather than relying only on cut-out clean speech regions. Our approach differs in that we start from a pretrained speaker verification extractor and adapt it with lightweight conditioning layers. For each selected local speaker, the local EEND output is converted into target-specific speaker-activity masks indicating silence, target-speaker activity, non-target activity, and overlap. These masks condition the embedding extractor frame by frame, allowing the model to exploit target-speaker evidence in both overlapped and non-overlapped regions while still producing a single embedding for the selected speaker.

We instantiate this idea with a pretrained WavLM-MHFA speaker embedding extractor~\cite{peng2023attention}, which combines a WavLM self-supervised speech encoder~\cite{chen2022wavlm} with multi-head factorized attentive (MHFA) pooling for speaker embedding extraction. Inspired by diarization-conditioned Whisper (DiCoW) for target-speaker ASR~\cite{polok2026dicow}, we inject speaker-activity information into intermediate WavLM representations using frame-level diarization-dependent transformations. The added conditioning layers are initialized to preserve the pretrained model behavior before adaptation, allowing us to start from a strong speaker verification model rather than training a diarization-conditioned extractor from scratch.

Our main contribution is a speaker-activity-conditioned WavLM-MHFA extractor, SAC-WavLM, that adapts a pretrained speaker verification model for target-speaker embedding extraction from full local EEND windows. We show that fine-tuning this model on a limited amount of real conversational data substantially reduces speaker-confusion errors and improves EEND-VC diarization performance, providing an effective alternative to large-scale simulated-mixture training. We also evaluate SAC-WavLM together with several modern embedding extractors using both agglomerative hierarchical clustering (AHC) and SphereVBx~\cite{palka2026SphereVBx} as clustering backends.

\section{Related Work}
\label{sec:related_work}

\textbf{Guided Speaker Embeddings.}
Horiguchi et al.~\cite{horiguchi2025guided} introduced guided speaker embeddings (GSE), which use target- and non-target-speaker activity as auxiliary information to extract a target-speaker embedding from multi-speaker audio. Speaker activity is provided to the embedding extractor together with the acoustic features and is also used to mask attentive pooling, allowing the model to use overlapped speech while suppressing regions where the target speaker is inactive. The model is trained on simulated three-speaker mixtures generated from VoxCeleb data using oracle speaker activities. A later extension~\cite{horiguchi2025mitigating} mitigates non-target-speaker bias by modifying modules that rely on global statistics, improving robustness in both low- and high-overlap conditions. These works are closest to ours in motivation, but they train dedicated guided embedding models on large-scale simulated mixtures, whereas we adapt a pretrained WavLM-MHFA speaker verification model using limited real conversational data.

\textbf{Diarization-Guided Multi-Speaker Embeddings.}
Kalda et al.~\cite{kalda2025diarization} extended guided embedding extraction to model multiple speakers jointly. Instead of extracting one target-speaker embedding per forward pass, their model uses information from a local segmentation model to produce embeddings for multiple active speakers at once, improving efficiency and matching the multi-speaker nature of diarization. This direction is complementary to ours: it addresses the efficiency and joint-modeling limitation of target-speaker embedding extraction, while our work focuses on adapting a strong pretrained extractor to conversational embedding extraction without relying on large-scale simulated-mixture training.

\textbf{Diarization-Conditioned Foundation Models.}
Our conditioning mechanism is inspired by diarization-conditioned Whisper (DiCoW)~\cite{polok2026dicow}, which injects diarization information into intermediate Transformer layers of Whisper~\cite{radford2023robust} for target-speaker ASR. We apply the same general idea to speaker embedding extraction: rather than changing the audio input, we condition intermediate representations of a pretrained model on speaker activity. This allows us to preserve the pretrained WavLM-MHFA initialization while adapting it to extract target-speaker embeddings from full local EEND windows.

\begin{figure}[t]
    \centering
    \includegraphics[width=\linewidth,trim=0.75cm 0.5cm 0.75cm 0.5cm,clip]{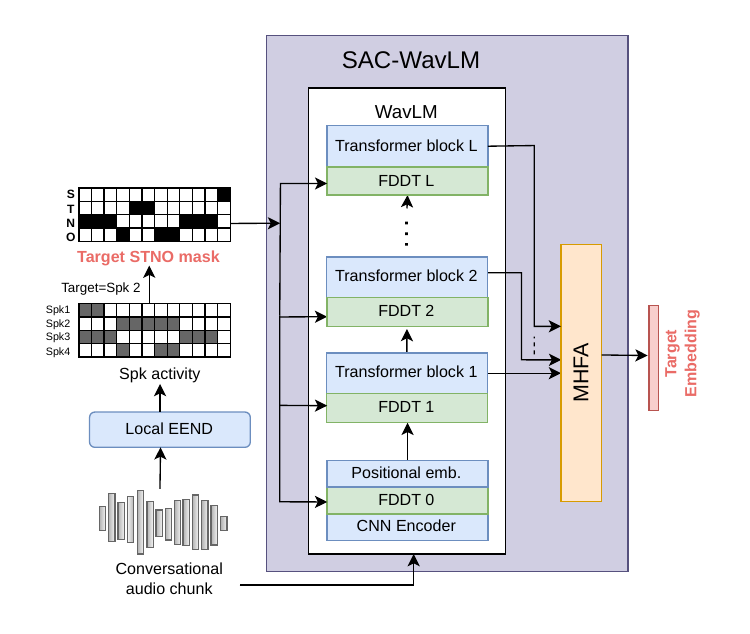}
    \caption{Block diagram of the proposed SAC-WavLM model. It shows extraction of a target embedding from a conversational audio chunk of the recording. STNO stands for \textbf{S}ilence, \textbf{T}arget, \textbf{N}on-target, \textbf{O}verlap.}
    \label{fig:block-diagram}
\end{figure}

\section{Proposed Method}
\label{sec:proposed_method}

Our goal is to adapt a pretrained speaker embedding extractor, originally trained on single-speaker speech segments, to extract an embedding for a selected speaker from a full local EEND window. The proposed speaker-activity-conditioned WavLM-MHFA model, denoted SAC-WavLM, keeps the architecture and initialization of this extractor, but adds speaker-activity conditioning through frame-level diarization-dependent transformation (FDDT) blocks inserted into the WavLM frontend. An overview of the model is shown in Fig.~\ref{fig:block-diagram}. Although this work instantiates the approach with WavLM-MHFA, the conditioning principle is not specific to WavLM and can be applied to other deep embedding extractors with intermediate frame-level representations.

\subsection{Pretrained WavLM-MHFA Extractor}

We build on the WavLM-MHFA speaker embedding extractor~\cite{peng2023attention}. This extractor combines a self-supervised pretrained WavLM frontend with an MHFA pooling backend, and the complete system is trained for speaker classification on single-speaker speech segments using the standard additive angular margin (AAM) loss. In Fig.~\ref{fig:block-diagram}, this baseline corresponds to the same WavLM-MHFA architecture without the STNO-mask input and without the FDDT blocks.

The extractor processes the raw waveform using the WavLM frontend, consisting of a convolutional feature encoder followed by a stack of Transformer layers. The multi-head factorized attentive pooling (MHFA) backend aggregates information across both time and WavLM layers to produce a speaker embedding. Let $\mathbf{Z}^l \in \mathbb{R}^{D \times F}$ denote the frame-level representation at the input of the $l$-th WavLM layer, where $D$ is the feature dimension and $F$ is the number of frames. MHFA first forms key and value representations by learned weighted combinations of WavLM layers:
\begin{equation}
\begin{aligned}
\mathbf{K} &= Linear_k\Big(\sum_{l=1}^L w_k^l\mathbf{Z}^l\Big), \\
\mathbf{V} &= Linear_v\Big(\sum_{l=1}^L w_v^l\mathbf{Z}^l\Big),
\end{aligned}
\label{eq:mhfa_kv}
\end{equation}
where $\mathbf{K},\mathbf{V}\in\mathbb{R}^{D' \times F}$, $w_k^l$ and $w_v^l$ are learnable layer weights, and $Linear_k$ and $Linear_v$ are linear projections. The attention weights over frames are computed using a learnable query matrix $\mathbf{Q}\in\mathbb{R}^{D'\times H}$, whose columns correspond to $H$ attention heads:
\begin{equation}
\mathbf{A} = \mathrm{softmax}(\mathbf{K}^T\mathbf{Q}),
\label{eq:mhfa_attention}
\end{equation}
where $\mathbf{A}\in\mathbb{R}^{F\times H}$. The value representations are pooled separately for each head and then concatenated to form the final embedding:
\begin{equation}
\mathbf{x} = Linear_{out}\big(\mathrm{vec}(\mathbf{V}\mathbf{A})\big),
\label{eq:mhfa_embedding}
\end{equation}
where $\mathrm{vec}(\cdot)$ denotes vectorization, i.e., concatenation of the head-specific pooled vectors. This layer-wise pooling motivates conditioning the model inside WavLM: transformations applied to intermediate WavLM representations can influence the representations that MHFA later selects and aggregates.

\subsection{Target-Specific STNO Masks}

Because SAC-WavLM processes the full local EEND window but must output an embedding for one selected local speaker, we convert the local speaker-activity labels into a target-specific STNO conditioning mask, following DiCoW~\cite{polok2026dicow}. For each selected target speaker, every frame is assigned to one of four mutually exclusive classes:
\begin{itemize}
    \item \textbf{S}ilence: no speaker is active,
    \item \textbf{T}arget: only the selected target speaker is active,
    \item \textbf{N}on-target: one or more non-target speakers are active and the target speaker is inactive,
    \item \textbf{O}verlap: the target speaker and at least one non-target speaker are active.
\end{itemize}

Thus, regardless of how many local speakers are present in the EEND window, SAC-WavLM always receives four mask streams. Their values are target-specific: selecting a different local speaker produces a different STNO mask for the same audio window. Let $S_t$, $T_t$, $N_t$, and $O_t$ denote the four STNO mask values at frame $t$. In this work, we use binary STNO masks, so exactly one of these values is one at each frame and the others are zero.

\subsection{Speaker-Activity-Conditioned WavLM}

Rather than modifying the waveform or concatenating speaker-activity features at the input, we inject the STNO conditioning into intermediate WavLM representations. This follows the DiCoW FDDT mechanism~\cite{polok2026dicow} and allows us to preserve the pretrained WavLM-MHFA extractor while adapting its internal representation flow for target-speaker embedding extraction.

For each selected WavLM Transformer layer $l$, an FDDT block applies one of four STNO-class-dependent affine transformations to each frame-level input representation. Let $\mathbf{z}^l_t \in \mathbb{R}^{D}$ denote the original input to the $l$-th Transformer layer at frame $t$. The FDDT block replaces it with the transformed representation
\begin{equation}
\begin{aligned}
\hat{\mathbf{z}}^l_t
&= (\mathbf{W}^l_S\mathbf{z}^l_t+\mathbf{b}^l_S) S_t
 + (\mathbf{W}^l_T\mathbf{z}^l_t+\mathbf{b}^l_T) T_t \\
&\quad + (\mathbf{W}^l_N\mathbf{z}^l_t+\mathbf{b}^l_N) N_t
 + (\mathbf{W}^l_O\mathbf{z}^l_t+\mathbf{b}^l_O) O_t ,
\end{aligned}
\label{eq:fddt}
\end{equation}
where the four affine transformations correspond to silence, target-speaker speech, non-target speech, and overlap, respectively. Since we use binary mutually exclusive STNO masks, Eq.~\eqref{eq:fddt} selects exactly one affine transformation for each frame. Following DiCoW, the transformation matrices are constrained to be diagonal in our implementation, keeping the number of additional parameters small.

The transformed representation $\hat{\mathbf{z}}^l_t$ is passed to the corresponding Transformer layer. After all WavLM layers have been processed, the MHFA backend aggregates the resulting layer representations and produces a single embedding for the selected target speaker. The same audio window can be processed repeatedly with different target-specific STNO masks to obtain embeddings for different local speakers.

\subsection{Initialization}
\label{sec:SACWavLM_Init}
A key design goal is to adapt the pretrained WavLM-MHFA extractor without disrupting its already useful speaker representation space. We therefore initialize SAC-WavLM from the baseline WavLM-MHFA model: all WavLM and MHFA parameters are copied from the pretrained extractor, and only the newly introduced FDDT parameters are initialized separately.

If all FDDT transformations are initialized as identities with zero biases, inserting the FDDT blocks leaves the pretrained extractor unchanged. Following DiCoW~\cite{polok2026dicow}, we use this identity initialization for all FDDT blocks except the first one.

The first FDDT block, denoted FDDT~$0$ in Fig.~\ref{fig:block-diagram}, uses suppressive initialization. Its silence and non-target transformations are initialized as scaled-down identities, reducing the acoustic information passed from frames that do not contain the target speaker, while target and overlap transformations are initialized as identities. The block is placed after the WavLM convolutional feature encoder but before positional embeddings are added, so that suppression affects acoustic features without modifying positional information. This gives the model a reasonable target-speaker behavior at initialization, while still allowing all transformations to be adapted during fine-tuning.

\subsection{Training Objective}
\label{sec:sac_wavlm_objective}
Speaker embedding extractors are commonly trained with the additive angular margin (AAM) loss. Let $\mathbf{x}_i$ be the embedding extracted from the $i$-th training example, let $p$ be its speaker label, and let $\mathbf{c}_k$ denote the learnable centroid corresponding to the speaker $k$. For the $i$-th training example, the AAM loss is defined as
\begin{equation}
\mathcal{L}_{\mathrm{AAM}}^{(i)} =
-\log
\frac{e^{s\cos(\theta_{ip}+m)}}
{e^{s\cos(\theta_{ip}+m)}+\sum_{k \neq p} e^{s\cos(\theta_{ik})}} ,
\label{eq:aam}
\end{equation}
where $\theta_{ik}$ is the angle between $\mathbf{x}_i$ and centroid $\mathbf{c}_k$, $s$ is the scale, and $m$ is the angular margin. In standard AAM training, the centroids $\mathbf{c}_k$ are learned jointly with the embedding extractor.

This standard setting is appropriate when many training speakers are available, as in VoxCeleb-based speaker verification training or training on simulated mixtures generated from VoxCeleb. However, when fine-tuning SAC-WavLM on real conversational datasets, only a limited number of speakers is available. Learning new centroids in this setting could encourage the model to overfit to those speakers instead of learning how to use the speaker-activity conditioning.

We therefore use fixed centroids when fine-tuning SAC-WavLM on real conversational data. For each training speaker, we compute a centroid by averaging embeddings extracted from clean single-speaker regions using the already trained embedding extractor. These centroids are then kept fixed during fine-tuning. This anchors SAC-WavLM to an existing speaker embedding space and reduces the risk of overfitting to the small real-data speaker set: 
the model is encouraged to produce embeddings compatible with the original extractor, while learning to use the full local window and the target-specific STNO conditioning.

\section{Experiments}

\begin{table}[t]
\caption{Training Data Statistics for Real and Simulated Datasets.}
\centering
\begin{tabular}{|l|cc|c|}
\hline
\textbf{Statistics\textbackslash{}Train Dataset} & \multicolumn{2}{c|}{\textbf{Real Dataset}} & {\textbf{Simulated Dataset}} \\ \hline
Source                    & AMI   &NSF                                                & VoxCeleb 2 dev \\
Total duration (hrs)                    & 80 &54                                                     & 1600                                                         \\ 
\#Total Speakers                        & 151 & 15                                                     & 5527                                                         \\
Recording duration (mins)               & 30-60 & 6-8                                                   & 2-55                                                         \\
\#Speakers per recording                & 3-5 & 4-8                                                      & 2-4                                                          \\ \hline
\end{tabular}
\label{tab:statistics}
\end{table}

We evaluate on standard conversational diarization benchmarks: AMI~\cite{carletta2005ami}, \mbox{AISHELL-4} (AIS)~\cite{fu21b_interspeech}, \mbox{AliMeeting} (AliM)~\cite{yu2022m2met}, \mbox{NOTSOFAR-1} (NSF)~\cite{vinnikov24_interspeech}, \mbox{MSDWild} (MSD)~\cite{liu22t_interspeech}, DIHARD3 full (DH3)~\cite{ryant21_interspeech}, and VoxConverse (VoxC)~\cite{chung20_interspeech}. For AMI and AliMeeting, we use the first channel of the far-field microphone array, while for NOTSOFAR-1, we follow the single-channel evaluation condition using beamformed recordings. We exclude RAMC~\cite{yang22h_interspeech} because we observed label noise in its training and test sets.

System performance is evaluated using diarization error rate (DER) without a forgiveness collar and considering overlapped speech using the dscore toolkit\footnote{\url{https://github.com/nryant/dscore}}. Hyperparameters are selected by minimizing DER on a compound development set, including the threshold and minimum cluster size for agglomerative hierarchical clustering (AHC), and the $F_A$ and $F_B$ parameters for SphereVBx~\cite{palka2026SphereVBx}.

\subsection{Baseline EEND-VC System}
\label{sec:Baseline}

All experiments use the DiariZen framework\footnote{\url{https://github.com/BUTSpeechFIT/DiariZen}} as an implementation of the EEND-VC pipeline described in Section~\ref{sec:intro}: a local EEND model predicts speaker activity in overlapping windows, embeddings are extracted for the local speakers, and clustering maps local speaker identities to global speaker identities. We keep the local EEND fixed and vary the speaker embedding extractor and/or clustering backend.

The recording is processed using $16\,\mathrm{s}$ local windows with a $1.6\,\mathrm{s}$ shift. The local EEND model is configured for up to four local speakers, all of which may be active in the same frame. For each local speaker, embeddings are extracted preferably from non-overlapped regions; overlapped regions are used only when no non-overlapped speech is available.

The baseline embedding extractor is the pretrained WavLM-MHFA model\footnote{\url{https://huggingface.co/JYP2024/WeSpeaker_SV_WavLM_MHFA}}, which consists of a WavLM Base+ encoder with 12 Transformer layers followed by an MHFA pooling backend with $64$ heads that produces 256-dimensional embeddings. The model is trained on the VoxCeleb2 development
set as discussed in \cite{peng2023attention}. 
Although DiariZen uses a relatively lightweight ResNet34~\cite{wang2023wespeaker} architecture with approximately 6.7M parameters, about $14 \times$ fewer than WavLM-MHFA, we adopt WavLM-MHFA because its Transformer-based architecture enables straightforward integration of speaker-activity conditioning while maintaining a comparable inference cost.
In addition to this baseline, we evaluate a WavLM-MHFA variant fine-tuned on the same real conversational data but without STNO conditioning or FDDT blocks referred to as WavLM-MHFA (Real), to separate the effect of domain fine-tuning from the proposed speaker-activity conditioning.

For clustering, we mainly use SphereVBx-PF~\cite{palka2026SphereVBx}, the parameter-free variant of SphereVBx. SphereVBx was recently shown to simplify and improve the DiariZen clustering stage by replacing the PLDA-based VBx backend with a spherical model better matched to modern length-normalized speaker embeddings. We also report results with AHC to compare with a more traditional clustering backend.

\subsection{SAC-WavLM Training Configuration}

During training, STNO masks are derived from ground-truth diarization labels. At evaluation time, we consider both oracle segmentation and the output of the DiariZen local EEND model, as described in the Results section.

\subsubsection{Training on Real Conversational Data}

To fine-tune SAC-WavLM on limited labeled conversational data, we use the training partitions of AMI and NOTSOFAR-1, which provide speaker identities that are consistent across recordings. We do not use the other evaluation corpora for training because they do not provide reliable speaker labels across sessions to construct global speaker identities. Together, AMI and NOTSOFAR-1 provide $134$ hours of training data and $166$ speakers, as summarized in Table~\ref{tab:statistics}.

For fixed-centroid AAM training described in Section~\ref{sec:sac_wavlm_objective}, we first compute one centroid for each training speaker. We extract $10\,\mathrm{s}$ embeddings from single-speaker regions using the pretrained WavLM-MHFA extractor and average embeddings belonging to the same speaker. These centroids are kept fixed during SAC-WavLM fine-tuning to reduce overfitting in the limited-speaker setting.

The SAC-WavLM model is initialized from the pretrained WavLM-MHFA extractor. All FDDT transformations are initialized as identities, except FDDT~$0$, where the silence and non-target transformations are scaled by $0.5$ as discussed in Section~\ref{sec:SACWavLM_Init}. Training is performed in two stages. First, only the FDDT and MHFA parameters are optimized for up to five epochs with a learning rate of $10^{-3}$. Then, the full model is fine-tuned for up to ten epochs using learning rates of $10^{-5}$ for the MHFA/FDDT backend and $2\times10^{-7}$ for WavLM. Each mini-batch contains 10 audio segments. For each segment, we use up to the four most talkative speakers, producing up to 40 STNO masks and corresponding speaker embeddings. The AAM scale is set to $s=14$, and the margin is set to zero based on validation performance.

\subsubsection{Training on Simulated Mixtures}

To compare fine-tuning on limited real conversational data with training on large-scale simulated speech, we also follow the simulation strategy of~\cite{landini2022simulated} to generate conversational mixtures from VoxCeleb2 development data. This gives approximately ten times more training audio and over thirty times more speakers than the real conversational training set; see Table~\ref{tab:statistics}.

The simulated-mixture variant is initialized in the same way as SAC-WavLM trained on real conversational data, starting from the pretrained WavLM-MHFA extractor with newly inserted FDDT blocks. Unlike fine-tuning on the limited real conversational data, simulated-mixture training uses standard AAM with learnable centroids. We first train only the centroids on non-overlapped single-speaker regions, using margin $m=0.2$, scale $s=32$, and a learning rate of $10^{-3}$ while keeping the rest of the model frozen. We then train on simulated multi-speaker mixtures: first, the FDDT and MHFA backend are optimized together with the centroids using a learning rate of $10^{-4}$; then, the full model is trained with learning rates of $10^{-5}$ for the backend and $10^{-6}$ for WavLM.

\subsubsection{Fine-Tuning After Simulated-Mixture Training}

We also evaluate a variant that is first trained on simulated mixtures and then fine-tuned on the real conversational data. After simulated-mixture training, we compute fixed centroids for the AMI and NOTSOFAR-1 training speakers using embeddings extracted by the simulated-mixture-trained model. These centroids are then frozen, and the full model is fine-tuned on real conversational data using learning rates of $10^{-6}$ for the MHFA/FDDT backend and $10^{-7}$ for WavLM, with scale $s=14$ and zero margin.

\begin{table*}[htbp]
\centering
\caption{Comparison of diarization speaker-confusion error (SC, \%) on benchmark datasets using SphereVBx clustering. FC denotes fixed centroids used during the final training stage. 
Relative improvements are computed with respect to the WavLM-MHFA baseline. \textbf{Bold} and \underline{underlined} values indicate the best and second-best results in each column.}
\label{tab:DER_oracle_diarizen}
\begin{tabular}{|l|c|c|cc|ccccc|cc|}
\hline
\textbf{Method} &  \textbf{FC} & \textbf{Segmentation}
& \textbf{NSF} & \textbf{AMI} & \textbf{AliM} & \textbf{MSDW}
& \textbf{DH3} & \textbf{AIS} & \textbf{VoxC}
& \textbf{\begin{tabular}[c]{@{}c@{}}Macro\\average\end{tabular}} & \textbf{\begin{tabular}[c]{@{}c@{}}Relative\\improv.\end{tabular}} \\
\hline
WavLM-MHFA (Baseline) &  \xmark & Oracle
& 4.42 & 2.03 & 1.11 & 2.08 & 2.74 & 2.65 & 1.91 & 2.42 & -- \\
WavLM-MHFA (Real)           &  \cmark & Oracle
& 4.97 & 1.38 & 1.06 & 2.48 & 2.95 & 2.24 & 1.77 & 2.41 & 0.5 \\
SAC-WavLM (Real)      &  \cmark & Oracle
& 2.09 & 1.58 & 0.28 & \underline{1.86} & 2.20 & \textbf{1.81} & 1.79 & 1.66 & 31.5 \\
SAC-WavLM (Sim.)      &  \xmark & Oracle
& \underline{1.62} & \underline{0.99} & \textbf{0.09} & 1.93 & \underline{2.17} & 2.38 & \underline{1.58} & \underline{1.54} & \underline{36.5} \\
SAC-WavLM (Sim. + Real) &  \cmark & Oracle
& \textbf{1.44} & \textbf{0.78} & \underline{0.13} & \textbf{1.58} & \textbf{2.11} & \underline{1.82} & \textbf{1.56} & \textbf{1.35} & \textbf{44.4} \\
\hline
WavLM-MHFA (Baseline) &  \xmark & DiariZen
& 7.23 & 3.53 & 1.95 & 4.94 & 2.88 & 2.73 & 2.29 & 3.65 & -- \\
WavLM-MHFA (Real) &  \cmark & DiariZen
& 7.74 & 2.71 & 2.06 & 5.11 & 2.93 & 2.56 & 2.18 & 3.61 & 1.0 \\
SAC-WavLM (Real)      &  \cmark & DiariZen
& 5.83 & \underline{2.20} & 1.34 & \textbf{5.04} & \underline{2.76} & \textbf{2.29} & \underline{2.08} & \textbf{3.08} & \textbf{15.7} \\
SAC-WavLM (Sim.)      &  \xmark & DiariZen
& \textbf{5.52} & 2.22 & \textbf{1.21} & 5.49 & \textbf{2.67} & 2.63 & 2.15 & 3.13 & 14.3 \\
SAC-WavLM (Sim. + Real) &  \cmark & DiariZen
& 6.05 & \textbf{2.08} & \textbf{1.21} & \underline{5.25} & 2.78 & \underline{2.49} & \textbf{2.07} & 3.13 & 14.3 \\
\hline
\end{tabular}
\end{table*}

\begin{table*}[htbp]
\centering
\caption{Comparison of DER (\%) with AHC and SphereVBx clustering on benchmark datasets using DiariZen segmentation. \textbf{Bold} and \underline{underlined} values indicate the best and second-best results in each column.}
\label{tab:DER_perform}
\begin{tabular}{|l|c|c|ccccccc|cc|}
\hline
\textbf{Method}
& \textbf{\begin{tabular}[c]{@{}c@{}}Diarization\\conditioning\end{tabular}}
& \textbf{Clustering}
& \textbf{AIS} & \textbf{AliM} & \textbf{AMI} & \textbf{DH3}
& \textbf{MSDW} & \textbf{NSF} & \textbf{VoxC}
& \textbf{\begin{tabular}[c]{@{}c@{}}Macro\\average\end{tabular}} & \textbf{\begin{tabular}[c]{@{}c@{}}Gap from\\oracle\end{tabular}} \\
\hline
\textit{Oracle clustering} & -- & --
& \textit{8.5} & \textit{9.9} & \textit{12.3} & \textit{12.4}
& \textit{13.4} & \textit{13.6} & \textit{7.6}
& \textit{11.1} & -- \\
\hline
WavLM-MHFA & \xmark & AHC
& 10.4 & 12.5 & 15.3 & 14.9 & 18.3 & 22.6 & 10.1 & 14.9 & 3.8 \\
ResNet34~\cite{wang2023wespeaker} & \xmark & AHC
& 9.9 & 12.1 & 14.6 & 15.6 & 18.7 & 21.3 & 10.1 & 14.6 & 3.5 \\
CAM++~\cite{wang2023cam++} & \xmark & AHC
& 10.6 & 12.1 & 15.4 & 14.8 & 18.1 & 20.2 & 10.0 & 14.4 & 3.3 \\
GSE (CAM++)~\cite{horiguchi2025mitigating} & \cmark & AHC
& 10.3 & 11.1 & 14.9 & 14.6 & 17.3 & 18.1 & 10.0 & \textbf{13.8} & 2.7 \\
SAC-WavLM (Real) & \cmark & AHC
& 10.2 & 11.6 & 14.7 & 14.7 & 17.2 & 18.8 & 9.8 & \underline{13.9} & 2.8 \\
\hline
WavLM-MHFA & \xmark & SphereVBx
& 10.4 & 10.8 & 14.5 & 14.1 & 15.9 & 17.3 & 9.3 & 13.2 & 2.1 \\
ResNet34~\cite{wang2023wespeaker} & \xmark & SphereVBx
& 9.8 & 10.7 & 13.5 & 14.1 & 15.4 & 16.8 & 9.3 & 12.8 & 1.7 \\
CAM++~\cite{wang2023cam++} & \xmark & SphereVBx
& 10.3 & 10.5 & 13.6 & 14.1 & 15.9 & 16.9 & 9.1 & 12.9 & 1.8 \\
GSE (CAM++)~\cite{horiguchi2025mitigating} & \cmark & SphereVBx
& \textbf{9.6} & \textbf{10.1} & \textbf{13.1} & \textbf{14.0} & \textbf{15.3} & \textbf{15.4} & \textbf{9.0} & \textbf{12.4} & 1.3 \\
SAC-WavLM (Real) & \cmark & SphereVBx
& 10.0 & 10.2 & 13.2 & \textbf{14.0} & 16.0 & 15.9 & 9.1 & \underline{12.6} & 1.5 \\
\hline
\end{tabular}
\end{table*}

\section{Results}

DER decomposes into missed speech, false alarm speech, and speaker-confusion (SC) errors. Since the local EEND output is fixed in our comparisons, changing the embedding extractor or clustering backend cannot recover missed speech or remove false alarms; it can only change speaker-confusion errors by altering how local speakers are linked across windows. We therefore first report SC to isolate the effect of the speaker embeddings and clustering. This is also important because missed-speech and false-alarm errors are strongly affected by annotation boundary conventions: conversational transcripts are often optimized for ASR and may contain loose speech boundaries~\cite{horiguchi2026tight}. In contrast, SC more directly measures whether speech from one speaker is assigned to another speaker.

Table~\ref{tab:DER_oracle_diarizen} reports SC across datasets for the WavLM-MHFA baseline and SAC-WavLM variants using SphereVBx clustering. The first half of the table uses oracle segmentation, where local speaker activity is taken from ground-truth labels. In this condition, missed-speech and false-alarm errors are zero, so DER equals SC. The table also reports macro-averages across datasets and relative improvements with respect to the WavLM-MHFA baseline. The AMI and NSF columns are especially relevant because these are the real conversational datasets used for fine-tuning.

The second row, WavLM-MHFA (Real), corresponds to fine-tuning the baseline WavLM-MHFA extractor on real conversational data without STNO conditioning; it gives almost no macro-average improvement over WavLM-MHFA (Baseline). This indicates that the improvement obtained with SAC-WavLM is not simply due to domain fine-tuning. In contrast, SAC-WavLM (Real), i.e., SAC-WavLM fine-tuned only on the limited real conversational data, reduces the macro-average SC from $2.42\%$ to $1.66\%$, corresponding to a $31.5\%$ relative improvement. Training on simulated mixtures also improves over the baseline, and SAC-WavLM (Sim. + Real), which is trained on simulated mixtures and then fine-tuned on real data, gives the best oracle-segmentation result, with a $44.4\%$ relative improvement. This suggests that simulated mixtures can provide useful additional training diversity when the conditioning masks are accurate.

The second half of Table~\ref{tab:DER_oracle_diarizen} reports results using estimated DiariZen local speaker activity. In this condition, the DiariZen output is used both as the local EEND decision in the EEND-VC pipeline and, for SAC-WavLM, to derive the target-specific STNO masks. SAC-WavLM (Real) gives the best macro-average SC, reducing it from $3.65\%$ to $3.08\%$ and giving a $15.7\%$ relative improvement. The simulated variants remain competitive, but they no longer improve over SAC-WavLM (Real). This suggests that fine-tuning directly on real conversational data better matches the estimated DiariZen outputs used at test time.

Table~\ref{tab:DER_perform} compares full DER across different embedding extractors and clustering backends using DiariZen local speaker activity. The oracle-clustering row gives a lower bound for each dataset by mapping DiariZen local speakers directly to ground-truth speaker labels, thereby removing clustering errors. The remaining gap from oracle clustering reflects the contribution of speaker-linking errors. All systems in this table use the same DiariZen local EEND output and the same clustering/post-processing pipeline; the main component being compared is the speaker embedding extractor.

The table includes both conventional speaker verification embeddings and diarization-conditioned embeddings. WavLM-MHFA is the baseline extractor used in our system. ResNet34~\cite{wang2023wespeaker} and CAM++~\cite{wang2023cam++} are additional single-speaker embedding extractors trained on VoxCeleb-style speaker verification data. GSE (CAM++)~\cite{horiguchi2025mitigating} is included as a previously published diarization-guided embedding extractor built on top of CAM++. This makes CAM++ the natural non-conditioned baseline for GSE, while WavLM-MHFA is the non-conditioned baseline for SAC-WavLM.

With AHC clustering, GSE (CAM++) gives the best macro-average DER in this comparison, reducing DER from $14.4\%$ for CAM++ to $13.8\%$. SAC-WavLM (Real) reduces DER from $14.9\%$ for WavLM-MHFA to $13.9\%$. Thus, both diarization-conditioned embedding extractors provide clear gains over their corresponding non-conditioned baselines. SAC-WavLM obtains these gains by adapting a pretrained WavLM-MHFA extractor using only limited real conversational data, whereas GSE relies on large-scale simulated-mixture training.

With SphereVBx clustering, all embedding extractors improve compared with AHC, confirming that the clustering backend remains an important part of the EEND-VC pipeline. GSE (CAM++) gives the best macro-average DER in this comparison, while SAC-WavLM (Real) gives the second-best result. Compared with their corresponding non-conditioned baselines, GSE reduces DER from $12.9\%$ to $12.4\%$, while SAC-WavLM reduces DER from $13.2\%$ to $12.6\%$. SAC-WavLM also reduces the gap from oracle clustering from $2.1\%$ to $1.5\%$. These results show that speaker-activity conditioning improves the embeddings used for linking local speakers, and that adapting a pretrained speaker verification model with limited real conversational data is an effective alternative to training a diarization-guided extractor on large simulated mixtures.

\begin{figure}[tb]
    \centering
    \includegraphics[width=0.5\textwidth,trim=0.7cm 2.2cm 0.1cm 1.8cm,clip]{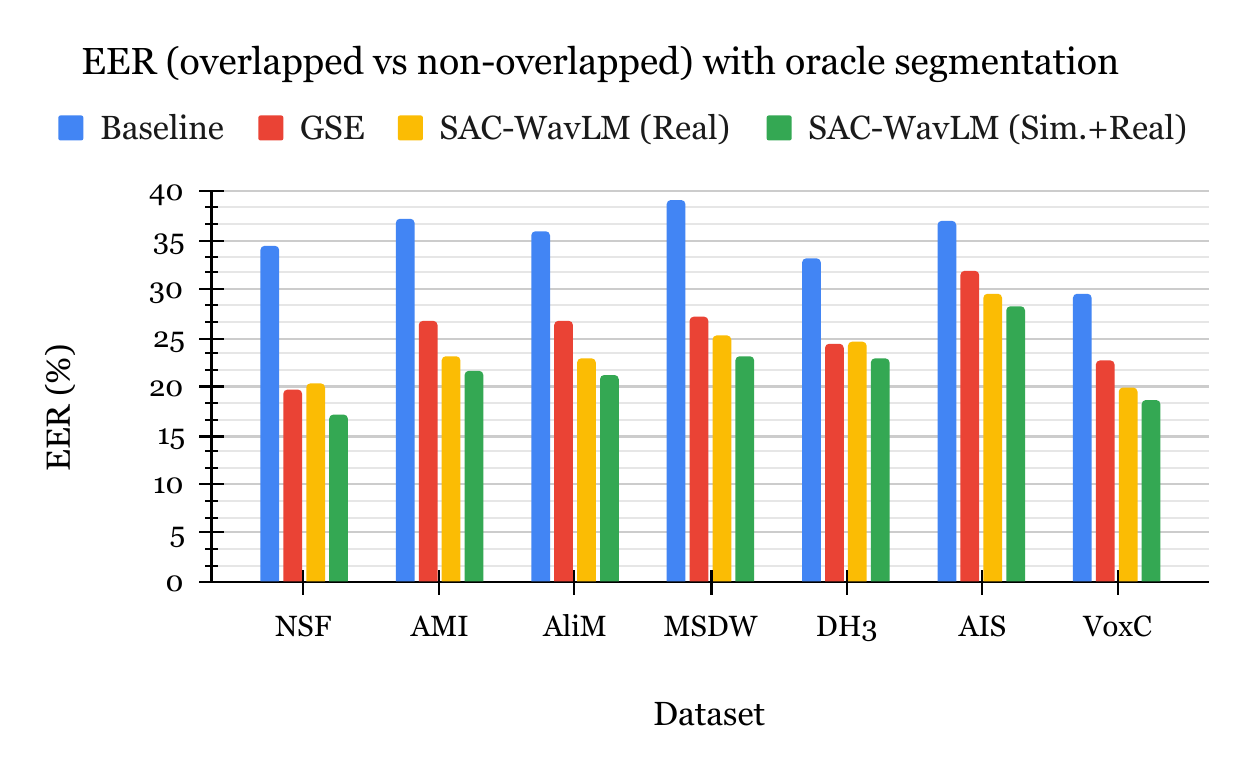}
    \caption{EER performance across datasets for overlapped ($<1.6\,\mathrm{s}$ target single speaker segments) vs non-overlapped ($\ge1.6\,\mathrm{s}$ target single speaker segments) speaker embeddings with ground-truth segmentation.}
    \label{fig:EER}
\end{figure}
\begin{figure}[tb]
    \centering
    \includegraphics[width=0.5\textwidth,trim=0.7cm 2.2cm 0.1cm 1.8cm,clip]{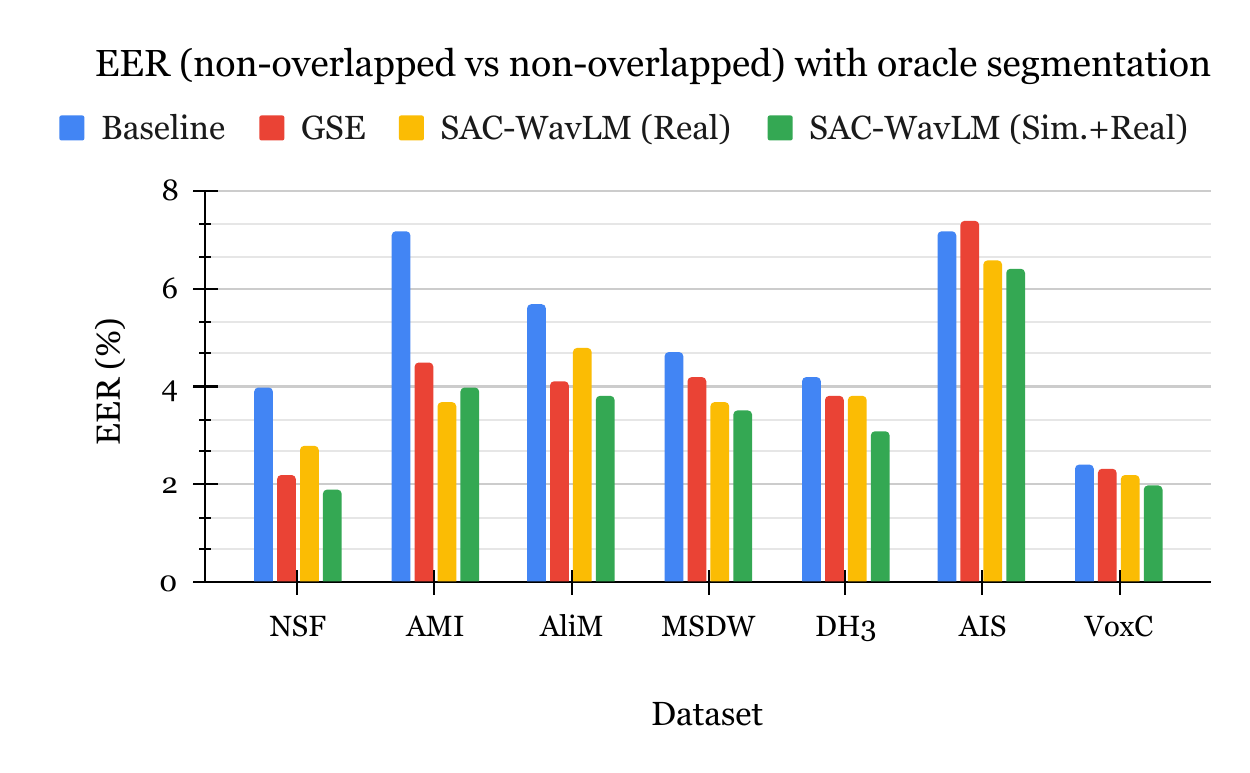}
    \caption{EER performance across datasets for pairs of non-overlapped (extracted from $\ge1.6\,\mathrm{s}$ target single speaker segments) speaker embeddings with ground-truth segmentation.}
    \label{fig:EER_clean}
\end{figure}

\section{Analysis}
To evaluate the discriminative quality of the extracted speaker embeddings independent of clustering, we compute the equal error rate (EER) using ground-truth speaker segmentation. Test-enrollment trials are formed within each recording to reflect the diarization setting. We consider two sets of verification trials.

Fig.~\ref{fig:EER} reports the EER computed using test embeddings extracted from overlapped speech segments containing less than $1.6\,\mathrm{s}$ of non-overlapped target speech. Enrollment embeddings are extracted from segments with at least $1.6\,\mathrm{s}$ of non-overlapped target speech. This evaluation measures the ability of the embedding extractor to associate challenging overlapped speech with the corresponding clean speaker representation. As expected, speaker-activity conditioning provides the largest improvements on datasets with higher overlap ratios. However, these gains do not translate proportionally to DER, since the downstream EEND-VC pipeline is also affected by inaccuracies in the local DiariZen speaker-activity estimates.

Fig.~\ref{fig:EER_clean} reports the EER using only non-overlapped speaker embeddings for both enrollment and test. The proposed method continues to outperform the baseline, indicating that speaker-activity conditioning improves speaker representations even for clean speech. We attribute this improvement to training on both overlapped and non-overlapped conversational regions, whereas the baseline is trained using only single-speaker segments. This advantage is particularly evident on datasets with substantial overlap, such as NSF, AMI, and AliMeeting.

\begin{figure}[t]
    \centering
    \includegraphics[width=\linewidth,trim=0.4cm 0.5cm 0.6cm 0.9cm,clip]{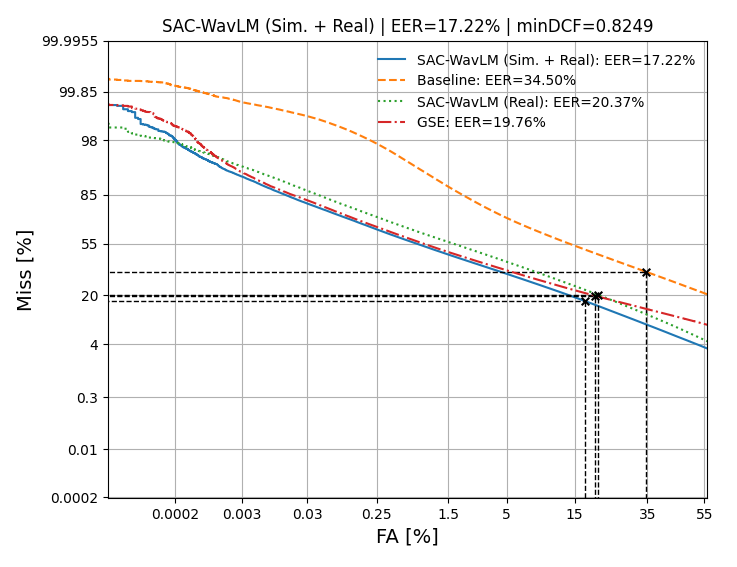}
    \caption{{Detection Error Tradeoff (DET) curve on NOTSOFAR-1 dataset. }}
    \label{fig:DET_curve_zoomed}
\end{figure}

Figure~\ref{fig:DET_curve_zoomed} presents the detection error tradeoff (DET) curves on the NOTSOFAR-1 test set for overlapped versus non-overlapped speaker verification trials. Unlike EER, which summarizes performance at a single operating point, the DET curves illustrate the trade-off between miss and false-alarm rates across different decision thresholds. The plot is zoomed around the EER region for clarity. The proposed speaker-activity-conditioned models consistently outperform the baseline across the operating range, demonstrating that the observed improvements are not limited to a specific threshold. Furthermore, the SAC-WavLM variants achieve performance comparable to, and occasionally better than, the GSE approach.

\section{Conclusion}

We proposed SAC-WavLM, a speaker-activity-conditioned embedding extractor for EEND-VC diarization. The method adapts a pretrained WavLM-MHFA speaker verification model to extract target-speaker embeddings from full local EEND windows, including both non-overlapped and overlapped speech. Inspired by DiCoW, speaker-activity information is injected into intermediate WavLM representations through frame-level diarization-dependent transformations, allowing the model to preserve the pretrained initialization while learning to use target-specific diarization cues.

Experiments across multiple conversational diarization benchmarks showed that SAC-WavLM substantially reduces speaker-confusion errors and improves full EEND-VC diarization performance. Fine-tuning on a limited amount of real conversational data was more robust with estimated DiariZen outputs than training on large simulated mixtures alone, and provided clear gains over the corresponding WavLM-MHFA baseline. We further compared several modern embedding extractors under common AHC and SphereVBx clustering backends, showing that speaker-activity-conditioned embeddings consistently improve speaker linking and that SphereVBx further reduces the remaining gap to oracle clustering.

\section*{Generative AI Use Disclosure}

Generative AI tools, such as ChatGPT, were used exclusively to refine language, paraphrasing, and improve manuscript readability. The scientific contributions, methodology, experiments, analysis, and conclusions were developed by the authors. The authors assume full responsibility for the accuracy, integrity and originality of the work and have approved its submission.
\IEEEtriggeratref{15}
\bibliographystyle{IEEEtran}
\bibliography{SLT2026_BibEntries.bib}

\end{document}